\documentclass[letterpaper]{JHEP3}
\usepackage{amsmath, amsfonts}
\usepackage{yfonts}
\usepackage{calc}
\usepackage[mathscr]{eucal}
\usepackage{epsfig}
\usepackage{array}
\usepackage{cite}

\def\be{\begin{equation}}
\def\ee{\end{equation}}

\def\href#1#2{#2}

\def\coeff#1#2{{\textstyle \frac {#1}{#2}}}

\newcommand{\tr}{\mathrm {tr}\,}
\newcommand{\mD}{m_{\rm D}}
\newcommand{\mtilde}{\widetilde m}
\newcommand{\gap}{m_{\rm gap}}
\newcommand{\E}{\mathbf E}
\renewcommand{\r}{\mathbf r}

\def\beq{\begin{equation}}
\def\eeq{\end{equation}}

\def\Im{\mop{Im}}

\def\Nfour{\mathcal{N}\,{=}\,4}

\def\Nc{N_{\rm c}}
\def\Nf{N_{\rm f}}

\def\Tc{T_{\rm c}}
\def\x{\mathbf {x}}
\def\P{\mathcal {P}}
\def\C{\mathcal {C}}
\def\T{\mathcal {T}}
\def\Im{\mathrm {Im}\,}
\def\rmax{r_{\rm max}}

\preprint{}

\title
    {%
    Debye screening in strongly coupled \boldmath $\Nfour$
    supersymmetric Yang-Mills plasma
    }

\author
    {%
    Dongsu~Bak,\!$^1$\footnote{\email{dsbak@mach.uos.ac.kr}}~
    Andreas~Karch,\!$^2$\footnote{\email{karch@phys.washington.edu}}~
    and
    Laurence~G.~Yaffe$^2$\footnote{\email{yaffe@phys.washington.edu}}
    \\
    $^1$Department of Physics, University of Seoul, Seoul 130-743, Korea
    \\
    $^2$Department of Physics, University of Washington, Seattle,
    WA 98195--1560, USA
    }

\abstract
    {
    Using the AdS/CFT correspondence,
    we examine the behavior of correlators of Polyakov loops
    and other operators
    in $\Nfour$ supersymmetric Yang-Mills theory
    at non-zero temperature.
    The implications for Debye screening in this
    strongly coupled non-Abelian plasma,
    and comparisons with available results for thermal QCD,
    are discussed.
    }


\begin{document}

\section{Introduction}

Debye screening is a characteristic feature of any plasma.
In Abelian plasmas, the electric field induced by a static test charge
decreases exponentially with distance from the charge,
$
    \langle \E(\r)\rangle \propto e^{-\mD |\r|}
$,
with $\mD$ denoting the Debye mass (or inverse Debye screening length).
The Debye mass cannot be defined so easily in
a non-Abelian plasma where the field strength
is only gauge-covariant, not gauge-invariant.%
\footnote
    {
    For example, in the presence of a test charge
    $\langle \tr \E(\r)^2 \rangle$
    is not proportional to $e^{-2\mD |\r|}$ at long distance.
    Rather, it falls exponentially as $e^{-|\r|/\xi}$,
    where $\xi$ is the longest correlation length of the system
    which, in weakly coupled plasmas, is due to static magnetic
    fluctuations.
    }
But, as discussed in Ref.~\cite{Arnold:1995bh},
the Debye mass can be given a precise, non-perturbative
definition (valid in both Abelian and non-Abelian theories)
as the smallest inverse correlation length in symmetry channels
which are odd under Euclidean time reflection.
In the Hilbert space interpretation of a theory,
Euclidean time reflection corresponds
to the product of charge conjugation
($\mathcal C$) and time reversal ($\mathcal T$).%
\footnote
    {
    This definition of the Debye mass is only applicable to theories which are
    invariant under $\cal CT$, in $\cal CT$ invariant
    equilibrium states.
    Hence this definition cannot be used with
    non-zero chemical potentials.
    See Ref.~\cite{Arnold:1995bh} for more discussion of this point.
    }

In a weakly coupled plasma,
the temperature $T$ sets the scale of
the energy or momentum of typical excitations
(gluons or quarks in a QCD plasma).
The Debye mass is parametrically smaller, $\mD \sim \sqrt \lambda T$,
with $\lambda \equiv g^2 \Nc$ the 't Hooft coupling
(defined at a scale on the order of $T$).
The Debye mass $\mD$ may be regarded as a thermally-induced effective mass
for static fluctuations in the (Euclidean) time component of the gauge field,
$A_0$.
The equilibrium behavior on distance scales large
compared to $\mD^{-1}$ may be described by a three-dimensional
effective theory which is obtained from the original four-dimensional
field theory (defined on a thermal circle of circumference $\beta = 1/T$)
by integrating out all non-static modes as well as the static component
of $A_0$.
What remains are just the static, spatial components of the gauge field,
whose dynamics are described by three-dimensional Yang-Mills theory%
\footnote
    {
    Up to higher dimensional irrelevant operators.
    Scalar fields, if present, generically receive thermally induced
    masses comparable to the Debye mass, and may also be integrated out.
    }
with a dimensionful 't~Hooft coupling $\lambda_3 = \lambda T$.
In non-Abelian theories,
fluctuations with wavenumbers on this scale are intrinsically non-perturbative.
Three-dimensional non-Abelian Yang-Mills theories are known to develop
a finite correlation length equal to the inverse of this scale
times a (non-perturbative) pure number.
Consequently, the mass gap $\gap$,
defined as the inverse of the longest correlation length,
of a weakly coupled non-Abelian plasma is of order $\lambda T$,
and is parametrically smaller than the $O(\sqrt\lambda T)$ Debye mass.

The leading weak coupling behavior of the Debye mass may be
calculated from one-loop thermal perturbation theory.
Subleading contributions to the Debye mass,
and the leading behavior of the mass gap,
depend on non-perturbative $\lambda T$ scale physics.
In this weak coupling regime, inverse correlation lengths in different
$\C\T$ (or Euclidean time reflection) odd symmetry channels differ
only by non-perturbative $O(\lambda T)$ amounts.
It is merely a convention to regard the Debye mass as the
smallest inverse correlation length in all $\C\T$-odd channels instead of using
the correlation length in a specific symmetry channel (such as, for example,
that of the imaginary part of the Polyakov loop).
As explained in Ref.~\cite{Arnold:1995bh}, regarding
the smallest $\C\T$-odd inverse correlation length as the Debye mass
leads to a particularly simple relation
between the next-to-leading order
contribution to $\mD$ and the expectation value of Wilson loops
in three-dimensional Yang-Mills theory.

The quark-gluon plasma (QGP) produced in heavy ion collision is,
for much of its evolution, thought to be rather strongly coupled
\cite{Shuryak:2004cy,Tannenbaum:2006ch},
and this has stimulated much theoretical interest in understanding
the dynamics of strongly coupled plasmas.
Although lattice simulations are a useful technique for
extracting static equilibrium quantities,
observables which are sensitive to real time evolution are
not generally accessible from Euclidean lattice simulations.%
\footnote
    {
    However, one can attempt to constrain suitable models
    of spectral functions by applying maximal entropy fitting methods
    to numerical data for Euclidean correlators.
    See, for example, Refs.~\cite{Asakawa:2000tr,Nakahara:1999vy,Aarts:2005zt,
    Aarts:2005hg,Wissel:2005pb,Datta:2004wc,Meyer:2007ic}.
    }

Gauge/string (or AdS/CFT) duality has provided new theoretical tools
for studying the dynamics of strongly coupled gauge theories and,
in particular, strongly coupled non-Abelian plasmas.
The most accessible system is maximally supersymmetric
Yang-Mills ($\Nfour$ SYM) theory in the limit of large $\Nc$
and large 't Hooft coupling.
Although this theory is manifestly not QCD,
at non-zero temperature $\Nfour$ SYM describes
a non-Abelian plasma composed of gauge bosons and
adjoint representation fermions and scalars
which shares many qualitative similarities with hot QCD.
Understanding the extent to which it is, or is not,
quantitatively similar to a QCD plasma is of
considerable current interest.

Several quantitative comparisons of dynamic properties of
$\Nfour$ SYM and QCD plasmas in the weakly coupled regime
are now available.
These include the shear viscosity \cite{Huot:2006ys},
the heavy quark energy loss ($dE/dx$) \cite{Chesler:2006gr},
and the photo-emission spectrum \cite{Caron-Huot:2006te}.
An interesting common feature has emerged from these weak-coupling
comparisons:
properties of $\Nfour$ SYM agree surprisingly well with those
of QCD provided one compares the two theories not at coinciding
values of the gauge coupling, but instead compares at coinciding
values of the Debye mass (or other closely related thermal scales).

One would like to understand the extent to which properties of
strongly coupled $\Nfour$ SYM plasma mimic those of a QCD plasma
at temperatures of perhaps 1.5--3 $T_c$ where the QCD plasma is
thought to be relatively strongly coupled.
In detail, this will undoubtedly depend on what observables
(or ratios of observables) are considered.

As one step in this direction, our goal in this paper is to
discuss what is known about the Debye mass and related phenomena
associated with screening in strongly coupled $\Nfour$ supersymmetric
Yang-Mills theory (via AdS/CFT duality) and to compare with
available data from lattice QCD.

Polyakov loop (or Wilson line) correlators provide the
simplest probe of screening in hot gauge theories.
A number of previous papers,
starting with Refs.~\cite{Rey:1998bq,Brandhuber:1998bs},
 have used AdS/CFT duality to examine
the behavior of Polyakov loop correlators in strongly coupled
$\Nfour$ SYM at non-zero temperature.
Confusing physical interpretations have appeared in several papers
(such as the suggestion that the connected part of the correlator
drops abruptly to zero at a finite separation \cite{Gubser:2006qh}), and
none of these papers specifically address the
non-perturbative definition of the Debye mass mentioned above.
Therefore, we revisit the analysis of Polyakov loop (and related)
correlators via AdS/CFT duality.
In particular, we emphasize that the connected parts of these correlators
must be non-vanishing at all separations and do have exponential long distance
behavior.%
\footnote
    {
    A closely related discussion with similar conclusions for the
    case of topologically trivial Wilson loop correlators
    at $T=0$ can be found in Ref.~\cite{Gross:1998gk}.
    }

We summarize what is known about the resulting correlation lengths
in different symmetry channels, and compare with available
data for corresponding correlation lengths obtained from
lattice simulations in QCD at temperatures above, but near,
the confinement/decon\-finement transition temperature.
An appendix discusses the transition from
weak to strong coupling in more detail, including partial results
on the first subleading $O(\lambda^{-3/2})$ strong coupling corrections
to the Debye mass.

It should be noted that some authors
(for example, Ref.~\cite{Kaczmarek:2005zn,Liu:2006nn})
refer to ``screening'' when discussing the behavior
of the static quark potential at non-asymptotic distances,
and interpret the phrase ``screening length'' to mean
some (loosely defined) notion of when the potential deviates
significantly from its zero temperature form.
This is not what we mean by screening.
We will always use ``Debye screening'' in the conventional sense,
in which a screening length characterizes
the asymptotic behavior of suitable correlators.

\section{Polyakov loop correlator from AdS/CFT}

For large $\Nc$ and large 't Hooft coupling $\lambda$, the
expectation values and correlation functions of Wilson loop
operators in the $\Nfour$ theory can be calculated using the
AdS/CFT correspondence.
The path integral in the five dimensional
gravity description in this limit is typically dominated by a semi-classical
open string worldsheet whose boundary coincides with the loop
\cite{Maldacena:1998im,Rey:1998ik}. In this framework, the
correlation function of two Polyakov loops,
\begin{equation}
    C(\x) \equiv \langle \P^*(\x) \P(0) \rangle \,,
\label{eq:full corr}
\end{equation}
was first obtained in
Refs.~\cite{Rey:1998bq,Brandhuber:1998bs}.  From this, one may extract
the static quark-antiquark potential (or more properly,
free energy) at finite temperature in the usual fashion,
\begin{equation}
    V_{q\bar q}(\x) \equiv -\beta^{-1} \ln \mathcal \, C(\x) \,.
\label{eq:Vqq}
\end{equation}

\begin{FIGURE}[ht]
{
\centering 
\includegraphics[scale=0.9,viewport=0 20 490 180,clip]{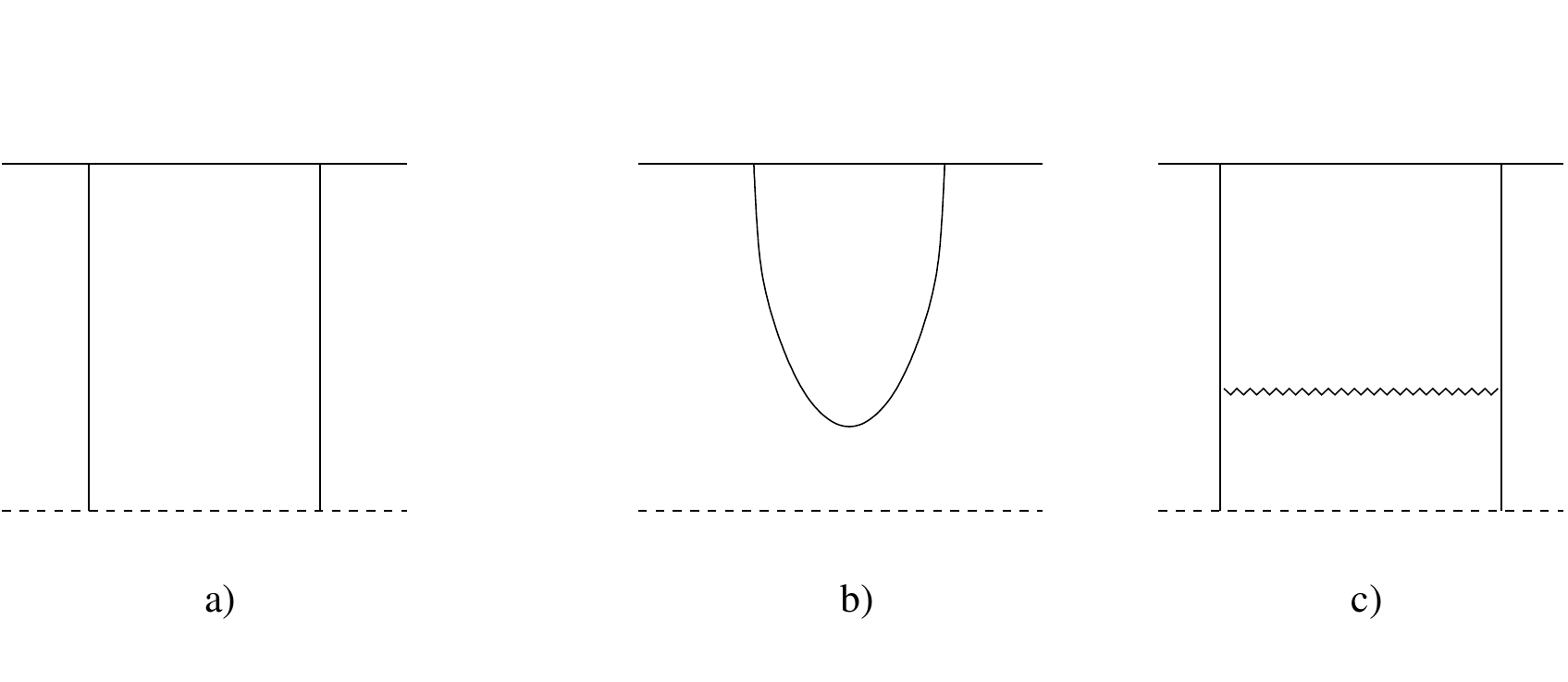}
\vspace*{-15pt}
\caption{\small
String configurations contributing to the Polyakov loop correlator.
The upper lines represent the boundary of the geometry
whereas the dashed lower lines
denote the location of the black hole horizon.
The periodic temporal direction is suppressed.
Configuration a) depicts a disconnected worldsheet,
with two independent strings running straight
from the boundary to the horizon.
This gives the disconnected part of the Polyakov loop correlator.
Configuration b) is a smooth connected worldsheet
which generates the dominant contribution to the connected correlator
for sufficiently small separations.
Configuration c)
is a connected contribution obtained by joining two otherwise
disconnected horizon-crossing string worldsheets by a small
tube representing the graviton propagator.
This gives the dominant contribution to the connected correlator
for sufficiently large separations.
\label{fig:strings}
}
}
\end{FIGURE}

In a non-confining phase (such as $\Nfour$ SYM at any $T > 0$),
the correlator (\ref{eq:full corr}) receives both connected and disconnected
contributions.
The disconnected contribution, depicted in Figure \ref{fig:strings}a,
corresponds to two straight strings stretching from the boundary to the
horizon.
For sufficiently small separations $|\mathbf x|$, the connected part of the
correlator is dominated by a smooth worldsheet of a string
connecting the two Polyakov loops.
This is illustrated in Fig.~\ref{fig:strings}b.
The regulated%
\footnote
    {
    To regulate the infinite worldsheet area,
    Refs.~\cite{Rey:1998bq,Brandhuber:1998bs} subtracted
    the on-shell action of two straight strings stretching from the
    horizon to the boundary.
    This corresponds to renormalizing Polyakov loops in such a way that
    their thermal expectation value is precisely $\Nc$.
    (We define the Polyakov loop $\P(\x)$ as a trace of the holonomy,
    not the trace divided by $\Nc$,
    so its expectation value in a deconfined phase is of order $\Nc$.)
    Strictly speaking, this prescription is not quite right,
    as it corresponds to a temperature dependent renormalization condition.
    Temperature independent renormalization
    (such as holographic renormalization \cite{Graham:1999pm}),
    would produce a result for the correlator (\ref {eq:full corr})
    which differs by an overall multiplicative factor.
    Such a finite multiplicative renormalization is irrelevant
    for our purposes, but its presence would complicate portions of the
    following discussion.
    Therefore, we will also adopt the simple subtraction scheme
    used in Ref.~\cite{Rey:1998bq,Brandhuber:1998bs}.
    \label{fn:renorm}
    }
action of this worldsheet is finite and leads to a quark-antiquark
free energy with attractive Coulombic short distance behavior.
This connected worldsheet configuration ceases to exist
(as a real solution)
beyond a critical separation $|\mathbf x| \ge \rmax =0.277/T$.
Even before that, at a separation $r_* \equiv 0.240 /T$,
the regulated action of the smooth connected
worldsheet (which is negative for $r < r_*$) crosses zero
and becomes positive for $r_* < |\x| < \rmax$.

As we discuss below, the connected Polyakov loop correlator,
\begin{equation}
    C_{\rm conn}(\x) \equiv
    \langle \P^*(\x) \P(0) \rangle - |\langle \P \rangle|^2 \,,
\label{eq:Cconn}
\end{equation}
must be a convex function which decreases monotonically with
increasing separation.
It cannot possibly drop to zero abruptly at some critical separation.
Since the smooth worldsheet configuration of Fig.~\ref{fig:strings}b ceases
to exist when $|\x|> \rmax$,
there must be some other configuration which yields the
dominant contribution to the connected correlator
in this regime.

To identify the missing contribution, it is helpful to recall the
appropriate $\Nc$ counting
(as was also explained in the closely related case of
correlation functions of two topologically trivial Wilson loops in
Ref.~\cite{Gross:1998gk}).
The disconnected part of the correlator $C(\x)$ scales as $\Nc^2$
when $\Nc \to \infty$.
Intuitively this factor of $\Nc^2$ can be seen in the supergravity
picture by noting that the endpoints of the string at the horizon
have a free Chan-Paton label indicating which D3 brane the string ends on
inside the horizon.
Hence there are really $\Nc^2$ configurations of two independent strings,
as depicted in Fig.~\ref{fig:strings}a.%
\footnote
    {
    Equivalently, in the Euclidean description of the AdS-Schwarschild
    background, where there is no horizon,
    each connected string worldsheet with $k$ holes and $l$ handles
    contributes at order $(\Nc)^{2-k-2l}$.
    So the two separate worldsheets,
    each with a single boundary hole and no handles, 
    in the disconnected contribution yield an overall $O(\Nc^2)$ result,
    while a connected worldsheet with two boundary holes is
    $O(\Nc^0)$.
    }

Large $\Nc$ factorization implies that the disconnected part of the
correlator $C(\x)$ is the only order $\Nc^2$ contribution;
the connected part is $O(\Nc^0)$, or smaller by a factor of $1/\Nc^2$.
Any connected worldsheet, such as the smooth
connected worldsheet found in Refs.~\cite{Rey:1998bq,Brandhuber:1998bs},
and sketched in Fig.~\ref{fig:strings}b,
does indeed produce an order one contribution to the correlator.
(Hence, it makes no sense to compare the
free energies of the string configurations shown in Fig.~\ref{fig:strings}a
and \ref{fig:strings}b, as they don't enter at the same order in $\Nc$.)

What then is the correct string
configuration which yields the leading connected contribution to the
Polyakov loop correlator at large separation?
The answer must be a connected worldsheet.
Since, for any $|\x|>\rmax$,
no such configuration exists solving the equations following from the
classical Nambu-Goto action, it must be a configuration with large
worldsheet curvatures so that quantum fluctuations of the worldsheet
become important and stabilize the configuration.
When $|\x| \sim \rmax$, it is difficult to find the resulting configuration
explicitly.
But when $|\x| \gg \rmax$,
the relevant worldsheet will approach a configuration of
two straight strings hanging
straight down toward the horizon, connected by a very thin tube,
as is illustrated in Fig~\ref{fig:strings}c.
This corresponds to the exchange of the lightest supergravity mode
coupling to both strings.

The emission and absorption vertices for supergravity modes each
come with a factor of the square root of
Newton's constant, $\sqrt{G_N} \sim {1}/{\Nc}$, so
the connected diagram with the supergravity exchange does indeed
contribute at the same $\Nc^0$ order as does the smooth connected worldsheet
in Fig.~\ref{fig:strings}b.
For large separations,
it is this ``connected by graviton exchange'' diagram
that gives the leading connected contribution to the Polyakov loop correlator.
The result will scale as $e^{- m|\x|}$, where
$m$ is the mass of the lightest supergravity
mode that can be sourced by the strings.%
\footnote
    {
    This is the same point which was made in Ref.~\cite{Gross:1998gk},
    where it was argued that Wilson loop correlators at large
    separation are dominated by the lightest supergravity mode.
    There was also an early attempt \cite{Danielsson:1999zt} along these lines
    to identify the Debye mass for the $\Nfour$ plasma.
    However, these authors considered the lightest mode in the
    spectrum which gives the mass gap instead of the Debye mass. In
    addition, at the time that paper was written the full supergravity
    fluctuation spectrum had not yet been worked out and it was
    incorrectly believed that the lightest mode would come from the
    dilaton. Only later was it realized that the lightest mode is
    actually part of the graviton.
    }

The real part of the Polyakov loop is even under $\C\T$
(or Euclidean time reflection),
while the imaginary part is $\C\T$-odd.%
\footnote
    {
    We have glossed over an irrelevant subtlety.
    Strictly speaking, the string configurations we have been
    discussing represent contributions to the correlators of
    Maldacena-Polyakov loops, which differ from ordinary Polyakov
    loops by the inclusion of a linear combination of scalar fields
    in the exponent \cite{Maldacena:1998im}.
    However, one may choose this linear combination of scalars to be
    even under both $\C$ and $\T$, so that the imaginary part of the loop
    corresponds to a $\C\T$-odd operator.
    It should also be noted that
    we are assuming that one is considering the pure phase
    of the deconfined $SU(\Nc)$ $\Nfour$ SYM plasma in which the expectation
    value of the Polyakov loop is real and positive.
    See Ref.~\cite{Arnold:1995bh} for more discussion of the significance
    of this point.
    }
As in the field theory discussion of Ref.~\cite{Arnold:1995bh},
when considering the correlator of just the imaginary part of
Polyakov loops,
\begin{equation}
    \widetilde C(\x)
    \equiv
    \langle \Im\P(\x) \, \Im\P(0) \rangle \,,
\end{equation}
the symmetries of the operator ${\rm Im}\, \P$ will restrict which
modes can be exchanged.
At large separations, the correlator $\widetilde C(\x)$
will be dominated by exchange of the
lightest mode in the supergravity multiplet
which can be sourced by the imaginary part of the Polyakov loop.
Let $\mtilde$ denote the mass of this mode.
Which $\C\T$-odd modes can contribute,
and the resulting value of $\mtilde$ will be discussed below.
Suitably distorted Polyakov loops can couple to all symmetry channels.
Therefore, what is clear on general grounds is that the Debye mass
(defined as the smallest $\C\T$-odd inverse correlation length)
is determined by the lightest $\C\T$-odd supergravity mode,
while the $\Im \P$ correlator has an exponential tail,
$ \widetilde C(\x) \sim e^{-\mtilde |\x|} $,
with a mass $\mtilde \ge \mD$.

To recap, the lightest mass of all supergravity
modes determines the mass gap $m_{\rm gap}$
(or inverse of the longest correlation length)
of the $\Nfour$ SYM plasma,
while it is the mass of the lightest $\C\T$-odd
supergravity mode which determines the Debye mass $\mD$.
We will see shortly that the lightest supergravity mode is $\C\T$ even,
and hence the Debye mass $\mD$ is different (and larger) than
the mass gap $m_{\rm gap}$.
Both of these masses will be
finite in the $\lambda \to\infty$ limit, and of order $T$.
As discussed in Ref.~\cite{Arnold:1995bh},
one can also consider a more refined classification
which distinguishes correlation lengths in all possible
symmetry channels.
In the strongly coupled $\Nfour$ SYM plasma,
these correlation
lengths will be determined by the lightest masses of supergravity
modes of a given symmetry.
This will be discussed explicitly below.

The full Polyakov loop correlator
(in the large $\Nc$ and large $\lambda$ regime)
is the sum of the disconnected,
smooth worldsheet,
and high curvature horizon-crossing worldsheet contributions
depicted in Fig.~\ref{fig:strings},
\begin{equation}
    C(\x) = C_a + C_b(\x) + C_c(\x) \,.
\label{eq:C_a}
\end{equation}
With the renormalization prescription
described in footnote \ref{fn:renorm},
the disconnected contribution has a fixed value,
\begin{equation}
    C_a = \Nc^2 \,.
\end{equation}
For $|\x| < r_*$,
the dominant connected contribution
comes from the smooth worldsheet and yields
\begin{equation}
    C_b(\x) = e^{ - \frac{\sqrt{\lambda}}{2 \pi} \mathcal A(|\x|)} \,,
\label{eq:C_b}
\end{equation}
where $\mathcal A(|\x|)$ is the regulated area of the smooth
worldsheet configuration of Refs.~\cite{Rey:1998bq,Brandhuber:1998bs},
measured in units of the AdS curvature radius.
The prefactor of $\sqrt{\lambda}/(2\pi)$
is the tension of the string in AdS units.
In this regime, the regulated area is negative,
so the connected correlator is exponentially large compared to unity.

For $|\x| > r_*$, the regulated area of the smooth worldsheet is positive
and its contribution is exponentially suppressed (in $\sqrt\lambda$).
The dominant contribution to the connected correlator, in this regime,
comes from the horizon-crossing string worldsheets involving thin tubes
with high curvature at their ends.
(A transition region whose width scales as $\lambda^{-1/2}$
connects these two regimes. This is discussed below.)
For $|\x| \gg r_*$,
this contribution to the correlator falls exponentially
with increasing separation,
\begin{equation}
    C_c(\x) \sim c \> e^{- m_{\rm gap} |\x|} \,,
\label{eq:C_c}
\end{equation}
where $m_{\rm gap}$ is the lightest mass of all supergravity modes.
The $\C\T$-odd correlator $\widetilde C(\x)$ will have faster
$e^{-\mtilde |\x|}$ fall-off for large separations,
while for small separations $|\x| < r_*$
it differs from $C_{\rm conn}(\x)/2$
only by exponentially small (in $\sqrt\lambda$) contributions
(arising from horizon-crossing worldsheets connected by a thin tube).

Although it is not strictly necessary for the subsequent portions
of this paper,
it is instructive to make a plot of the Polyakov loop correlator,
and the resulting quark-antiquark potential (\ref{eq:Vqq}),
for various values of $\lambda$.
Fig.~\ref{fig:Cconn} shows the connected part of the correlator
for $\lambda = 10$, $10^2$, $10^3$ and $10^4$,
while Fig.~\ref{fig:Vstatic} shows the resulting static quark-antiquark
potential, given by%
\footnote
    {
    Note that,
    because the static quark-antiquark potential involves the
    logarithm of the full correlator and not just its connected part,
    $V_{q\bar q}(\x)$
    is {\em not}\/ simply proportional to the regulated area $\mathcal A(\x)$
    when $r < r_*$.
    Only when $\mathcal A(|\x|) \ll -4\pi (\ln \Nc)/\sqrt\lambda$
    does the static potential reduce to
    $\beta^{-1} \frac{\sqrt{\lambda}}{2 \pi} \mathcal A(|\x|)$.
    }%
$^,$%
\footnote
    {
    These plots are qualitatively, and even semi-quantitatively, correct.
    But they are not (and cannot be) exact for several reasons.
    First, only the asymptotic form (\ref{eq:C_c}) of $C_c(\x)$ is known.
    As we discuss in the next section, the value of the mass gap
    can be extracted from known results for linearized gravitational
    fluctuations in the AdS-Schwarschild background
    (and one finds $m_{\rm gap} = 2.3361 \, \pi T$).
    But neither the $O(1)$ amplitude $c$, nor corrections
    to the asymptotic form, are known.
    To make a qualitatively correct figure,
    we have assumed that $C_c(\x)$ is a pure exponential
    at all distances, with a value of unity at $r_{\rm max}$.
    Second, the regulated smooth worldsheet area $\mathcal A(\x)$ is only
    defined (as a real function) for $|\x| < r_{\rm max}$.
    For $r_* < |\x| \le r_{\rm max}$, the contribution
    $C_b(\x)$ is a nonperturbative, exponentially small correction
    to the correlator.
    Nevertheless, when plotting the correlator for any fixed finite
    value of $\lambda$, one must decide how to treat this contribution
    when $|\x|$ passes $r_{\rm max}$.
    Our lack of knowledge about the right answer (which might be
    computable using suitable analytic continuation in the string
    worldsheet functional integral) amounts to a non-perturbative
    uncertainty which, formally, is much less important than
    the (unknown) corrections to the dominant contribution $C_c(\x)$
    which are suppressed by inverse powers of $\lambda$.
    But if one regards $C_b(\x)$ as abruptly dropping to zero
    when $|\x|$ exceeds $r_{\rm max}$ then, for any finite $\lambda$,
    this introduces a discontinuity in the correlator (and a delta-function
    spike in the force $F_{q\bar q}(\x) \equiv -\mathbf\nabla V_{q\bar q}(\x)$)
    which is obviously unphysical.
    To avoid such spurious artifacts (and solely for the purpose of making
    a qualitatively correct plot), we made an ad-hoc but smooth extrapolation
    of $\mathcal A(\x)$ beyond $r_{\rm max}$.
    Specifically, for $|x| > r_{\rm max}$
    we used an exponential function of the form
    $
	C_b(\x) = \alpha + \beta \, e^{-\gamma|\x|}
    $
    with $\alpha$, $\beta$ and $\gamma$ determined by matching
    the value and first two radial derivatives of $C_b(\x)$ at $r_{\rm max}$.
    }
\begin{equation}
    V_{q\bar q}(\x) = -\beta^{-1} \ln [C_a + C_b(\x) + C_c(\x)] \,,
\end{equation}
for the same set of $\lambda$ values.
As these plots clearly illustrate, as $\lambda$ increases the
correlator, and the static potential, develop a kink at $r = r_*$,
with exponential sensitivity to $\lambda$ at smaller separations
and almost no sensitivity to $\lambda$ at larger separations.

\begin{FIGURE}[ht]
{
\centering 
\includegraphics[scale=0.7]{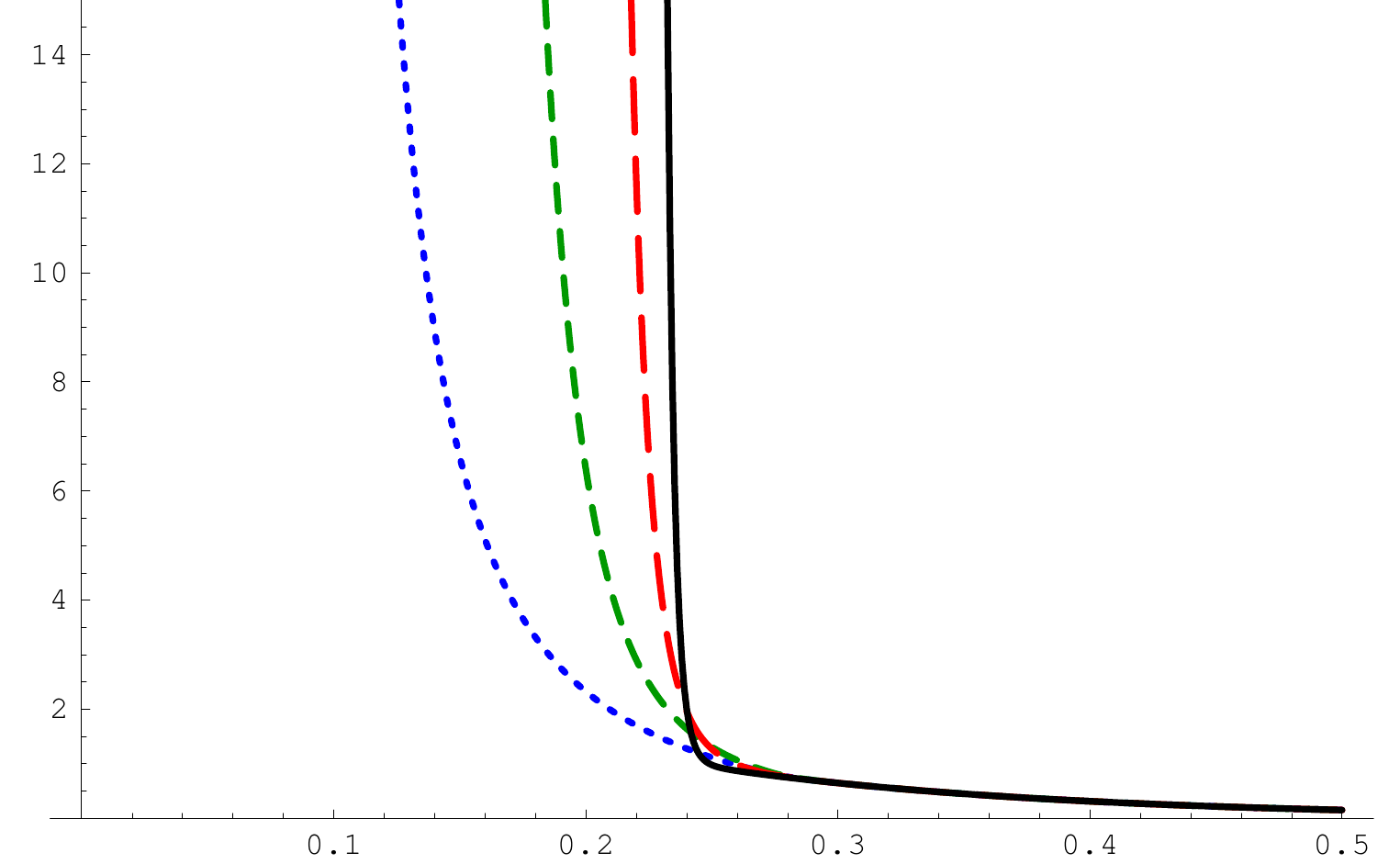}
\begin{picture}(0,0)(0,0)
\put(5,8){$|\x|\,T$}
\put(-315,200){$C_{\rm conn}(\x)$}
\end{picture}
\caption
    {\small
    The connected part of the Polyakov loop correlator $C_{\rm conn}(\x)$
    plotted as a function of the separation $|\x|$,
    for $\lambda = 10$ (blue, dotted),
    $10^2$ (green, short dash),
    $10^3$ (red, long dash), and
    $10^4$ (black, solid).
    As $\lambda\to\infty$, the correlator develops a kink at
    $|\x| = r_* = 0.240/T$.
    At separations larger than $r_*$, the exponential decay of the
    correlator is nearly independent of $\lambda$.
    \label{fig:Cconn}
    }
}
\end{FIGURE}

\begin{FIGURE}[ht]
{
\centering 
\includegraphics[scale=0.7]{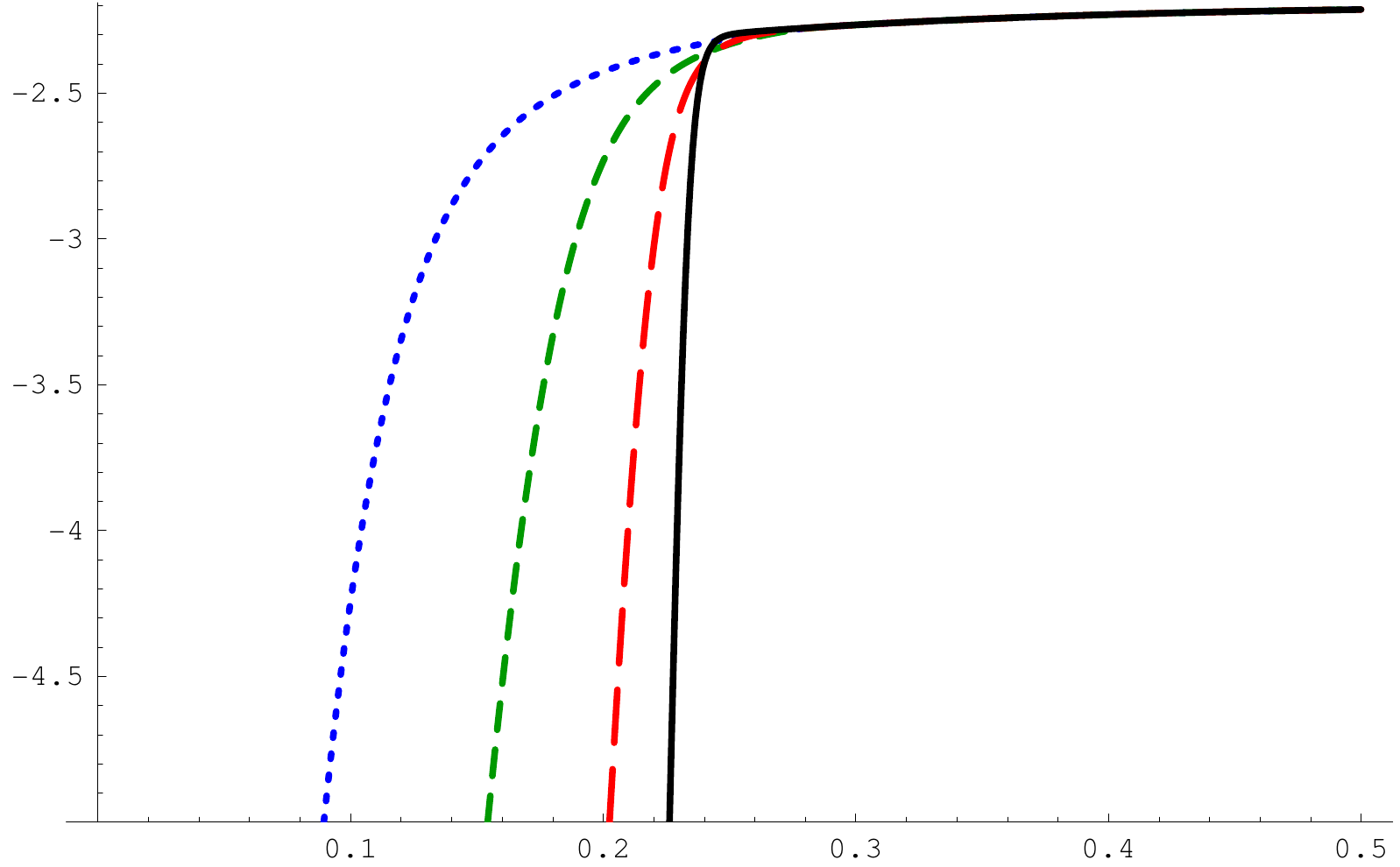}
\begin{picture}(0,0)(0,0)
\put(5,8){$|\x|\, T$}
\put(-320,210){$V_{q\bar q}(\x)/T$}
\end{picture}
\caption{\small
    The static quark-antiquark potential (or free-energy)
    $V_{q\bar q}(\x)$ plotted as a function of the separation $|\x|$,
    for $\Nc = 3$ and
    $\lambda = 10$ (blue, dotted),
    $10^2$ (green, short dash),
    $10^3$ (red, long dash), and
    $10^4$ (black, solid).
    As $\lambda\to\infty$, the static potential develops a kink at
    $|\x| = r_* = 0.240/T$.
    At separations larger than $r_*$, the potential continues
    to rise in a manner nearly independent of $\lambda$,
    and asymptotically approaches $-T \ln \Nc^2$.
    \label{fig:Vstatic}
    }
}
\end{FIGURE}

If one considers the scaling with $\lambda$ of
the static quark-antiquark potential, or the force
$\mathrm F_{q\bar q}(\x) \equiv -\mathrm \nabla V_{q\bar q}(\x)$
between a static quark and antiquark,
then (outside the narrow transition region)
for $r < r_*$ the force is $O(\sqrt\lambda)$
while for $r > r_*$ the force is $O(1)$.
Focusing only on this comparison might lead one to think that
the $r > r_*$ contribution should be regarded as
a subleading $1/\lambda$-suppressed correction to the
leading large $\lambda$ contribution from the smooth worldsheet.
This perspective is, however, highly misleading.
The smooth worldsheet (Fig.~\ref{fig:strings}b) and
the high-curvature horizon-crossing worldsheet (Fig.~\ref{fig:strings}c)
represent different stationary points of the worldsheet effective action.
Except within the narrow transition region,
one configuration gives the leading large $\lambda$ contribution
to the connected correlator $C_{\rm conn}(\x)$, while the other
gives an exponentially suppressed relative contribution.
But which configuration is leading and which is exponentially suppressed
switches at the cross-over separation $r_*$.
Obviously, the contribution of Fig.~\ref{fig:strings}c has nothing
to do with $1/\lambda$ corrections which will arise from small fluctuations
around the smooth string worldsheet of Fig.~\ref{fig:strings}b.

As a final point of this section, let us clarify the nature of
the cross-over near $r_*$.
In any thermal field theory,
consider the Euclidean two point correlator,
\be
    {\cal G} (\x) = \langle W^{\dagger} (\x) W(0)\rangle \,,
\ee
of any operator $W$, where $\x$ is a spatial separation
between the two operators.
Finite temperature implies that one direction ---
which one normally regards as (Euclidean) time ---
is compactified.
But one may equally well choose to regard the compactified direction
as a spatial direction, and view the separation $\x$ as
defining the Euclidean time direction.
A Hilbert space interpretation based on this definition
of time immediately yields the spectral representation,
\be
    {\cal G}(\x) =\sum_n e^{-\epsilon_n |\x|} \, |c_n|^2\,,
\label{eq:spec-rep}
\ee
for the correlator.
Here $\{ \epsilon_n \}$ are the excitation energies of
eigenstates $\{ |n\rangle \}$ of the Hamiltonian
(for the spatially compactified system)
and $c_n \equiv \langle n | W |0\rangle$.
Since every term in the sum is positive,
this representation shows that the correlator
$\mathcal G(\x)$, as a function of separation $|\x|$,
is positive, monotonically decreasing, and convex.
Moreover, if the connected part of the correlator,
$\mathcal G(\x) - \mathcal G(\infty)$, is non-zero at some
separation $\x$, then it must be non-zero (and positive)
at all $\x$.
In other words, a connected correlator cannot be non-zero
for some range of separations, and then vanish identically
beyond a critical separation.

These general principles are completely consistent with
the above description of the Polyakov loop correlator $C(\x)$.
But the change in behavior near $r_*$
implies that two very different energy scales
contribute to the spectral representation for $C(\x)$.
Specifically, there are states with $O(T)$ energies starting
at $m_{\rm gap}$,
and in addition states with $O(\sqrt\lambda T)$ energies
starting at a threshold of $\eta\sqrt\lambda/(2\pi)$,
where $\eta = 20.7 \, T$ is the slope with which the
regulated area of the smooth worldsheet crosses zero
at $|\x| = r_*$,
$
    \mathcal A(|\x|) \sim \eta (|\x| {-} r_*)
$.
For any finite value of the 't Hooft coupling $\lambda$,
the correlator will be a smooth convex function of separation.
The low energy states completely dominate when $|\x|$ is big compared to $r_*$
while the high energy states dominate for $|\x|$ well below $r_*$.
High and low energy states make comparable contributions
only in a narrow cross-over region around $r_*$, whose
relative width $\delta r/r_*$ scales as $\lambda^{-1/2}$.
In the $\lambda\to\infty$ limit, this width shrinks to zero
and the correlator develops a kink, but remains positive,
monotonic, and convex.

\section{Correlation lengths from supergravity modes}

Fortunately the analysis of the complete IIB supergravity spectrum,
in $R$-symmetry singlet channels,
on the AdS black hole background has already been performed in
Ref.~\cite{Brower:2000rp} (following earlier work on portions
of the spectrum in Ref.~\cite{Csaki:1998qr}).
In those papers the spectrum was presented
as glueball masses of a three-dimensional confining gauge theory
(containing Kaluza-Klein towers whose spacing is comparable
to the mass of the lightest glueballs).
We find it much more natural to interpret the supergravity spectrum as
giving us information about the finite temperature $\Nfour$ SYM plasma.
As just discussed, the supergravity spectrum directly yields the
spatial correlation lengths, in different symmetry channels,
of the hot plasma.

\begin{TABLE}[ht]
{
\renewcommand{\arraystretch}{1.2}
\begin{tabular*}{140mm}{@{\extracolsep\fill}||c||c|c||c||c||}
\hline \hline
SUGRA mode
& \phantom{abc}$J^{PCT}$\phantom{abc}
& \phantom{abc}${\mathscr J}^{CR_t}_{R_y}$\phantom{abc}\vphantom{\bigg|}
& \phantom{abcd}$m_0$\phantom{abcd}
&SYM operator\\
\hline \hline
$G_{00}$ & $0^{+++}$ & $0^{++}_+$ & 2.3361 & $T_{00}$\\
\hline
$G_{ij}$ & $2^{+++}$ & $2^{++}$ & 3.4041 & $T_{ij}$\\
\hline
$a$ & $0^{-+-}$ & $0^{+-}_-$ & 3.4041 & ${\rm tr} \, E \cdot B$\\
\hline
$\phi$ & $0^{+++}$ & $0^{++}_+$ & 3.4041 & $ {\cal L} $\\
\hline
$G_{i0}$ & $1^{++-}$ & $1^{+-}$ & 4.3217 & $T_{i0}$\\
\hline
$B_{ij}$ & $1^{+--}$ & $0^{-+}_-$ &  5.1085  & $\mathcal O_{ij}$ \\
\hline
$C_{ij}$ & $1^{--+}$ & $0^{--}_+$ &  5.1085  & $\mathcal O_{30}$ \\
\hline
$B_{i0}$ & $1^{--+}$ & $1^{--}$ &  6.6537  & $\mathcal O_{i0}$ \\
\hline
$C_{i0}$ & $1^{+--}$ & $1^{-+}$ &  6.6537   & $\mathcal O_{3j}$ \\
\hline
$G^{a}_{a}$ & $0^{+++}$ & $0^{++}_+$ & 7.4116  & ${\rm tr} \, F^4$  \\
\hline
\end{tabular*}
\caption
    {Spectrum of $R$-singlet IIB supergravity modes in units of $\pi T$,
    together with SYM operators dual to the indicated mode
    (see text for details).
    $J^{PCT}$ are quantum numbers for $(3{+}1)$-dimensional Poincar\'e
    symmetry,
    while ${\mathscr J}_{R_y}^{CR_t}$ are quantum numbers for the
    $O(2) \times Z_2$ Euclidean symmetry in the $xyt$-plane
    transverse to the $x_3$ direction (plus charge conjugation).
    }
\label{tab:classIIB}
}
\end{TABLE}

For the reader's convenience, we reproduce
in Table~\ref{tab:classIIB}
the results for the
supergravity modes from Ref.~\cite{Brower:2000rp}.
The relevant supergravity fields are the metric
$G_{\mu \nu}$, the dilaton-axion pair $e^{- \phi} + ia$, and the
NSNS and RR two forms $B_{\mu \nu}$ and $C_{\mu \nu}$.
For each one of these fields,
the mass%
\footnote
    {
    This terminology is a bit sloppy.
    The mass $m_0$ is the smallest magnitude of an imaginary wavevector
    $\vec\kappa$ for which the linearized supergravity equations
    for the various fields
    (in the Euclidean AdS black hole geometry)
    admit solutions whose only spacetime dependence is of the form
    $\exp (\vec\kappa \cdot \x) f(u)$, where $u$ is the radial AdS coordinate.
    This is precisely what is needed to determine the correlation lengths
    of Euclidean correlators.
    }
of the lightest mode with a given polarization is listed,
in units of $\pi T$,
together with its transformation properties $J^{PCT}$
under $(3{+}1)$-dimensional Poincar\'e symmetry.
In this table (as in Ref.~\cite{Brower:2000rp}),
the longitudinal direction defined by
the spatial wave vector $\bf k$ of the mode is taken to point in the
$x_3$ direction.
For our application to two-point correlators,
this corresponds to the direction of the spatial separation $\x$ of
the operators. The indices $i$ and $j$ denote transverse spatial
directions (corresponding to $x_1$ and $x_2$),
and $a$ refers to the directions along the internal $S^5$.
The table also lists the quantum numbers $\mathscr J^{CR_t}_{R_y}$
for the
$O(2) \times Z_2$ Euclidean symmetry in the $x,y,t$ plane
transverse to the longitudinal direction, plus charge conjugation,
which is the relevant symmetry group for a transfer matrix
in the $x_3$ direction.
Here, $\mathscr J$ is the two-dimensional angular momentum,
$R_t$ is the eigenvalue corresponding to the reflection $t \to -t$,
$R_y$ is the eigenvalue for the reflection $y \to -y$,
and $C$, as usual, is the eigenvalue for charge conjugation.
(Note that only one-dimensional irreducible representations
with $\mathscr J = 0$ need to be labeled with $R_y$.
Two dimensional representations with $\mathscr J \ne 0$ contain
components with both signs of $R_y$.)
As mentioned in the Introduction,
Euclidean time reflection is the product of
time reversal and charge conjugation, so $R_t = CT$.

The field $B_{\mu\nu}$ couples to the current $J_F^{\mu\nu}$
of the fundamental string which is electrically charged,
while the $C_{\mu\nu}$ field couples to the magnetically charged
D-string current.
Hence these fields are both charge conjugation odd,
but have opposite parity and time reversal assignments.

Using the standard dictionary of the AdS/CFT correspondence, all
operators dual to the above supergravity modes can be identified.
In particular,
the operator dual to the graviton corresponds to
the energy-momentum tensor ${T}_{\mu\nu}$ of $\Nfour$
super Yang-Mills theory.%
\footnote
    {
    The analysis of Ref.~\cite{Brower:2000rp} used a gauge choice
    in which the independent modes of the metric and two-form fluctuations
    correspond to polarizations without any indices in the longitudinal ($x_3$)
    or AdS radial directions.
    Hence the absence of these field components in the table.
    Because $T_{\mu\nu}$ is conserved,
    the corresponding ``missing'' components of the energy-momentum tensor
    ($T_{03}$, $T_{i3}$ and $T_{33}$) are not independent from the
    components shown in the table.
    }
The axion is dual to the Pontryagin density
$
    \coeff 14 \, {\rm tr} F^{\mu\nu}\widetilde F_{\mu\nu}
    =
    {\rm tr} \, E \cdot B
$,
whereas the
dilaton is dual to
the $\Nfour$ Lagrange
density $\mathcal L = \coeff 14 \, {\rm tr}\, F^{\mu\nu} F_{\mu\nu} + \cdots$.
Since the $\C$ and $\T$ quantum
numbers of $E$ and $B$ are $-+$ and $--$, respectively,
one finds $+-$ for $E\cdot B$, and $++$ for ${\cal L}$,
in agreement with the listed assignments for the
axion and dilaton modes.
The operator dual to the lowest mode of $G^a_a$
involves the trace of four powers of the field strength;
a more explicit form may be found in Ref.~\cite{D'Hoker:2002aw}.
The two-form fields $B_{\mu\nu}$ and $C_{\mu\nu}$ couple
to the SYM antisymmetric tensor operator
${\cal O}_{\mu\nu}$ \cite{Das:1998ei},
which is the $\Nfour$ completion
of the QCD operator
\be
O_{\mu\nu} \equiv
{\rm tr} \left[  F_{\mu\alpha} F^{\alpha\beta} F_{\nu\beta}+
\coeff{1}{ 4} \,
F_{\mu\nu} F^{\alpha\beta} F_{\alpha\beta} \right]
\,.
\ee
Five dimensional antisymmetric tensor fields have in general
six real independent on-shell degrees of freedom.
But in type IIB supergravity on AdS$_5\times S^5$,
there is an extra constraint linearly
relating $B_{\mu\nu}$ and $C_{\mu\nu}$ \cite{Brower:2000rp}.
Hence the total on-shell degrees of freedom in these two fields
are reduced to only six.
Among these, $B_{ij}$ is dual to
${\cal O}_{ij}$ and $C_{ij}$ to $\epsilon_{ij03} \, {\cal O}_{03}$.
One can also find that
$B_{i0}$ is dual to ${\cal O}_{i0}$
while $C_{i0}$ is dual to $\epsilon_{i03j} \, {\cal O}_{3j}$.
Since $B_{\mu\nu}+i C_{\mu\nu}$ or
$e^{-\phi}+ia$
are  multiplets of the $SL(2,R)$
symmetry of the tree-level IIB supergravity,
their spectrum should be degenerate. Indeed this can  be checked
from Table~\ref{tab:classIIB}.
Note that the different correlation lengths
for $O_{12}$ and $O_{3j}$
reflect our choice of $x_3$ as the longitudinal direction.

Examining Table~\ref{tab:classIIB},
one sees that
the true mass gap of the $\Nfour$ SYM plasma (in the $\lambda\to\infty$
and $\Nc\to\infty$ limits) arises from the
time-time component of the graviton.
This belongs to the $\C\T$ even sector,
as asserted earlier.
The SYM mass gap is thus
\be
m_{\rm gap} = 2.3361 \ (\pi T) \,,
\label{eq:mgap}
\ee
and this characterizes the longest correlations
in this non-Abelian plasma.
On the other hand, the lowest mass in a $\C\T$-odd channel comes from
the response of the axion field $a$. Therefore the Debye mass for the
 $\Nfour$ SYM plasma in the strong coupling limit is given by
\be
\mD = 3.4041 \ (\pi T) \,.
\label{eq:mD}
\ee
The imaginary part of the Polyakov loop has $J^{PCT} = 0^{+-+}$
or $\mathscr J^{CR_t}_{R_y} = 0^{--}_+$
and, in particular, is $\C$-odd.
Hence exchange of the $\C$-even (but $\T$-odd) axion field
cannot contribute to the $\Im \P$ correlator.
Moreover, as Table~\ref{tab:classIIB} shows,
there is no supergravity field with $0^{+-+}$ quantum numbers.
However, the three-form field strength
$(dC)_{123}$ does have precisely this symmetry,
and its spectrum is the same as that of $C_{12}$ for our choice of
longitudinal direction $x_3$.
Consequently, exchange of the two-form $C$ can contribute to the
$\Im \P$ correlator, implying that%
\footnote
    {
    An open string does not source the two-form $C$ at tree level,
    but fermion fluctuations on the string do couple to the
    RR field-strength.
    This implies that the $e^{-\widetilde m |\x|}$ tail in the
    $\widetilde C(\x)$ correlator will have a coefficient
    suppressed by $1/\lambda$ relative to the short distance
    contributions.
    This shifts the center of the cross-over region by
    a tiny relative amount of order $O[(\ln\lambda)/\lambda]$.
    }
\be
\mtilde = 5.1085 \ (\pi T) \,.
\label{eq:mtilde}
\ee

Note that the supergravity analysis predicts not only the values of
various correlation lengths, it also identifies which local operators
in the dual field theory couple most directly to the lightest modes
(in the strong coupling regime).
In principle, it should be possible to confirm the predictions
(\ref{eq:mgap})--(\ref{eq:mtilde}) via lattice simulations of
the relevant correlators
in $\Nfour$ SYM --- but this is not yet practically feasible.

\section{Comparison to QCD}

Data for screening masses ({\em i.e.}, inverse correlation lengths)
in various symmetry channels,
in $SU(2)$ and $SU(3)$ pure gauge theory, are available from
$4d$ lattice simulations
\cite{Datta:2002je,Datta:1999yu,Datta:1998eb,Datta:1998em},
as well as $3d$ simulations using the dimensionally
reduced high temperature effective theory
\cite{Hart:2000ha,Philipsen:2000qv,Hart:1999dj,Laine:1999hh,Kajantie:1997pd}.
Results from both approaches are generally consistent with each other
at the 10--15\% level down to $T = 2\Tc$.
(For a good summary of the status of lattice calculations of
thermal correlation lengths, see Ref.~\cite{Laermann:2003cv}.)
The dimensional reduction approach allows one to treat
(with no additional computational complexity)
QCD with any number of massless fermions.
Using this approach, Ref.~\cite{Hart:2000ha} reports results for
many different symmetry channels and zero, two, three or four quark flavors.

In Yang-Mills theory, the ratios of screening masses to the temperature are
very nearly temperature independent for $1.5 \Tc \le T \lesssim 4T$,
but (at least in the $0^{+++}$ and $0^{+-+}$ channels)
decrease substantially as the temperature drops below 1.5$\Tc$ and approaches
the confinement transition
\cite{Datta:1999yu,Datta:2002je}.
The degree of similarity between thermal QCD and $\Nfour$ SYM is
surely greatest within the window, $1.5 \Tc \le T \lesssim 4T$,
where screening masses scale linearly with temperature.
Comparison of results for screening masses in $SU(2)$ and $SU(3)$
Yang-Mills theory reveal very little dependence on $\Nc$ if
$\lambda = g^2\Nc$ is held fixed.
Consequently, the difference between $\Nc = 3$ and $\Nc = \infty$
is expected to be quite small \cite{Hart:2000ha}.

To make specific comparisons, we will use the results from
Ref.~\cite{Hart:2000ha}
for $T = 2 \Tc$,%
\footnote
    {
    Actually, these results are for $T = 2\Lambda_{\overline{MS}}$.
    The ratio of $\Tc/\Lambda_{\overline{MS}}$ is close to one,
    but is not known with very high accuracy when $\Nf \ne 0$.
    }
and we will also focus on the case of $\Nf = 2$,
where simulations were performed at a finer lattice spacing than
for other (non-zero) values of $\Nf$.
QCD lattice simulations find that the smallest
thermal screening mass is in the scalar $0^{+++}$ channel,
in agreement with the strong coupling $\Nfour$ SYM analysis.
Table \ref{tab:compare} shows results from Ref.~\cite{Hart:2000ha}
for the ratios of screening masses
in different symmetry channels
relative to the mass gap
(the lightest screening mass)
for $\Nf = 2$ QCD at $T = 2\Tc$,
together with the corresponding $\lambda = \infty$ $\Nfour$ SYM
results from Table~\ref{tab:classIIB}.%
\footnote
    {
    In contrast to the (lighter) spin 0 and spin 2 channels,
    Ref.~\cite{Hart:2000ha} specified
    the time reflection symmetry but not the charge conjugation
    properties of the spin one operators they used.
    We have assumed that their spin one results should be compared
    with the lightest $\mathscr J = 1$ channels with the given value
    of $R_t$ which, for strongly coupled $\Nfour$ SYM,
    are the spin one channels indicated in Table~\ref{tab:compare}
    and Fig.~\ref{fig:compare}.
    }
The same information is displayed graphically in Fig.~\ref{fig:compare}.

\begin{TABLE}[t]
{
\renewcommand{\arraystretch}{1.2}
\begin{tabular}{||c||c|c||c||}
\hline \hline
\vphantom{\bigg|}
~~~$\mathscr J^{CR_t}_{R_y}$~ &
$\Nf\,{=}\,2$ QCD & $\Nfour$ SYM & SUGRA mode
\\
\hline\hline
$0^{+-}_-$ & 1.45(4) & 1.46 & $a$\\\hline
$0^{--}_+$ & 1.79(4) & 2.19 & $C_{ij}$ \\\hline
$2^{++}  $ & 2.05(6) & 1.46 & $G_{ij}$ \\\hline
$0^{-+}_-$ & 2.12(7) & 2.19 & $B_{ij}$ \\\hline
$1^{+-}  $ & 2.31(10) & 1.85 & $G_{i0}$ \\\hline
$1^{-+}  $ & 2.79(12) & 2.85 & $C_{i0}$ \\
\hline
\end{tabular}
\vspace*{15pt}
\caption
    {
    Ratios of screening masses in indicated symmetry channels
    to the mass gap, for $\Nf = 2$ QCD at $2 \Tc$,
    and $\lambda = \infty, \Nc= \infty$ $\Nfour$ SYM.
    For both QCD and strongly coupled $\Nfour$ SYM,
    the mass gap is the screening mass in the $0^{++}_+$ channel.
    \label{tab:compare}
    }
}
\end{TABLE}

\begin{FIGURE}[t]
{
\newdimen\ymin
\newdimen\ymax
\newdimen\dy
\def\errbar(#1,#2,#3,#4,#5)%
    {%
    \put(#1,#2){\makebox(0,0){$\square$}}
    }%
\setlength{\unitlength}{40pt}
\begin{picture}(7,4.0)(-.5,-.5)
\put(.1,0){\line(1,0){6.5}}
\put(.1,1){\line(1,0){.1}}
\put(.1,2){\line(1,0){.1}}
\put(.1,3){\line(1,0){.1}}
\put(1,0){\line(0,1){.1}}
\put(2,0){\line(0,1){.1}}
\put(3,0){\line(0,1){.1}}
\put(4,0){\line(0,1){.1}}
\put(5,0){\line(0,1){.1}}
\put(6,0){\line(0,1){.1}}
\put(0,1){\makebox(0,0)[r]{1}}
\put(0,2){\makebox(0,0)[r]{2}}
\put(0,3){\makebox(0,0)[r]{3}}
\put(.1,0){\line(0,1){3.3}}
\put(-.7,1.9){\makebox(0,0){$\displaystyle \frac m{m_{\rm gap}}$}}
\put(.40,-.35){\makebox(0,0){$\mathscr J^{CR_t}_{R_y}$\,:}}
\put(1.15,-.35){\makebox(0,0){$0^{+-}_-$}}
\put(2.15,-.35){\makebox(0,0){$0^{--}_+$}}
\put(3.15,-.35){\makebox(0,0){$2^{++}_{\vphantom+}$}}
\put(4.15,-.35){\makebox(0,0){$0^{-+}_-$}}
\put(5.15,-.35){\makebox(0,0){$1^{+-}_{\vphantom+}$}}
\put(6.15,-.35){\makebox(0,0){$1^{-+}_{\vphantom+}$}}
\errbar(1,1.447,0.0382,1.428,1.466)
\errbar(2,1.791,0.0393,1.771,1.811)
\errbar(3,2.052,0.0600,2.022,2.082)
\errbar(4,2.119,0.0605,2.090,2.150)
\errbar(5,2.307,0.0957,2.259,2.355)
\errbar(6,2.794,0.1192,2.734,2.854)
\put(1,1.46){\makebox(0,0){$\Diamond$}}
\put(2,2.19){\makebox(0,0){$\Diamond$}}
\put(3,1.46){\makebox(0,0){$\Diamond$}}
\put(4,2.19){\makebox(0,0){$\Diamond$}}
\put(5,1.85){\makebox(0,0){$\Diamond$}}
\put(6,2.85){\makebox(0,0){$\Diamond$}}
\end{picture}
\vspace*{15pt}
\caption
    {
    Ratios of screening masses in indicated symmetry channels
    to the mass gap (the $0^{+++}$ screening mass),
    for $\Nf = 2$ QCD at $2 \Tc$ (squares),
    and $\lambda = \infty$ $\Nfour$ SYM (diamonds).
    The reported statistical errors on the QCD lattice results
    are smaller than size of the squares.
    \label{fig:compare}
    }
}
\end{FIGURE}

The lightest $\C\T$ (or $R_t$) odd screening mass,
which defines the Debye mass,
is in the $\mathscr J^{CR_t}_{R_y} = 0^{+-}_-$ (axion) channel
in both QCD and strongly coupled $\Nfour$ SYM.
As show in the above table and figure,
there is essentially perfect agreement between the two theories
for the value of the Debye mass to mass gap ratio,
$\mD/m_{\rm gap} \approx 1.46$.
This remarkable agreement (to better than a percent)
is surely somewhat fortuitous;
in other channels
the discrepancy between QCD and $\Nfour$ SYM for these
screening mass to mass gap ratios varies between
a few percent and 30\% (for the $2^{++}$ channel).
Overall, however, there is rather good agreement between
the two theories.

Instead of focusing on ratios of screening masses,
if one looks at their absolute size (in units of the temperature),
then one finds that screening masses in QCD (at $T \approx 2T_{\rm c}$)
\cite{Hart:2000ha}
are significantly smaller than in $\lambda = \infty$ $\Nfour$ SYM:
\begin{eqnarray}
    \frac {m_{\rm gap}}{\pi T}
    &=&
    \begin{cases}
    1.25(2) \,, & \Nf\,{=}\,2 \mbox { QCD, } T\,{=}\,2T_{\rm c};
    \\
    2.34 \,, & \Nfour \mbox{ SYM, } \lambda\,{=}\,\infty,
    \end{cases}
\\
    \frac {\mD}{\pi T}
    &=&
    \begin{cases}
    1.80(4) \,, & \Nf\,{=}\,2 \mbox { QCD, } T\,{=}\,2T_{\rm c};
    \\
    3.40 \,, & \Nfour \mbox{ SYM, } \lambda\,{=}\,\infty.
    \end{cases}
\label{eq:mDcompare}
\end{eqnarray}

In $\Nfour$ SYM, these masses (divided by $T$) are
dimensionless functions depending only on the coupling $\lambda$.
In the weak coupling regime, $m_{\rm gap}/T$ vanishes linearly
as $\lambda \to 0$, while $\mD/T$ scales as $\sqrt\lambda$.
Presumably, both masses grow monotonically with increasing $\lambda$
and asymptote to the above $\lambda=\infty$, $\Nfour$ SYM values.

The smaller values for the Debye mass (and the mass gap) in
QCD at $2 \Tc$, as compared to $\lambda=\infty$ SYM values,
naturally suggests that QCD plasma, in this temperature range,
is most similar to $\Nfour$ SYM plasma at an intermediate value
of $\lambda$ which is neither in the asymptotically weak,
nor asymptotically strong coupling regimes.
To test this hypothesis quantitatively, it will be necessary to
compute subleading (in $1/\lambda$) corrections to $\lambda = \infty$
screening masses in $\Nfour$ SYM plasma.
The Appendix discusses this in more detail, and reports some partial results
on the first subleading correction to $\mD$.

Other possible extensions of this work
include a comparison of correlation lengths
associated with flavor current correlators probing the dynamics of
fundamental representation matter added to $\Nfour$ SYM.
In the gravitational dual, this corresponds to the
addition of D7 flavor branes as described in Ref.~\cite{Karch:2002sh}.
Lattice data is now available for a variety of mesonic correlation
lengths in hot QCD
\cite{Wissel:2005pb,Pushkina:2004wa,Laermann:2003cv,Laermann:2001vg}.
We hope that this, and similar work, will clarify the degree to which
$\Nfour$ supersymmetric Yang-Mills plasma can mimic quantitative
properties of the quark-gluon plasma produced in heavy ion collisions.

\section*{Acknowledgments}
D.B. would like to thank the Particle Theory Group of
the University of Washington for the warm  hospitality.
L.G.Y thanks the Galileo Galilei Institute for Theoretical
Physics for its hospitality, and the INFN for partial support during the
completion of this work.
We also thank Ofer Aharony and Michael Gutperle
for helpful conversations and communications.
This work was supported in part by the U.S. Department
of Energy under Grant No.~DE-FG02-96ER40956 and by KOSEF
SRC CQUeST R11-2005-021.

\newpage
\begin{appendix}

\section{From weak to strong coupling}

The leading weak-coupling behavior of the Debye mass in $\Nfour$ SYM
is determined by the one-loop thermal correction to the gluon
self-energy. One easily finds the lowest-order result
\cite{Vazquez-Mozo:1999ic,Kim:1999sg} \be
    \mD^{(0)} = {\sqrt{2 \lambda}} \ T \,.
\ee
This is larger than the corresponding result for massless QCD
(with $\Nc = 3$)
by a factor
of $\sqrt 6 \approx 2.45$ for $\Nf = 0$,
$3/\sqrt 2 \approx 2.12$ for $\Nf = 2$,
or 2 for $\Nf = 3$
(with $\Nf$ the number of quark flavors).

The next-to-leading order weak-coupling correction to $\mD$
contains a logarithmic term whose coefficient may be easily extracted using
perturbative effective field theory methods,
together with a non-perturbative $O(\lambda T)$ contribution
which can only be computed via numerical lattice simulations.
The analysis of Ref.~\cite{Arnold:1995bh} shows that
changing the matter content of the theory from QCD
to that of $\Nfour$ SYM does not affect the next-to-leading order
correction to the Debye mass
(other than changing the value of $\mD^{(0)}$).
The result is
\be
    \mD
    =
    \mD^{(0)}
    +
    \frac {\lambda T}{4\pi} \, \ln\frac{\mD^{(0)}}{\lambda T}
    +
   \kappa \, \lambda T
    +
    O(\lambda^{3/2} T)
    ,
\label{eq:mDweak}
\ee
with the non-perturbative coefficient
$\kappa = 0.64(2)$ for $\Nc=3$
\cite{Laine:1999hh}.%
\footnote
    {
    For $SU(2)$, $\kappa = 0.63(2)$ \cite{Laine:1999hh}.
    Determinations of $\kappa$ from lattice simulations with
    $\Nc > 3$ have, to our knowledge, not yet been performed,
    but the dependence of $\kappa$ on $\Nc$ is clearly very weak.
    Comparisons with four-dimensional simulations also show that,
    in pure Yang-Mills,
    $O(\lambda^{3/2} T)$ and higher contributions are not large
    compared to the $O(\lambda T)$ terms when
    $T \gtrsim 2 T_{\rm c}$ \cite{Laine:1999hh}.
    }
The weak-coupling expansion of the Debye mass is much better
behaved in $\Nfour$ SYM than in QCD
due to the larger value of $\mD^{(0)}$.
(The $O(\lambda T)$ terms only become larger than
$\mD^{(0)}$ when $\lambda$ exceeds 5.6.)

The strong-coupling results for inverse
correlation lengths derived in this paper have corrections
which are suppressed by inverse powers of the 't Hooft coupling $\lambda$,
arising from $\alpha'$ corrections to type IIB supergravity.
The first subleading corrections for the dilaton
were investigated in Ref.\cite{Csaki:1998qr}, with the result
\be
m_\phi = 3.4041 \, \big[ 1-  1.39\, \zeta(3) \,
\lambda^{-3/2} + \cdots \big] (\pi T)\,.
\ee

To find the analogous subleading correction for the Debye mass,
one cannot invoke the $SL(2,R)$ symmetry of the tree-level IIB supergravity;
an independent calculation of the axion mode is required.
Corrections to the axion mass come
from the original IIB action expanded around the $\alpha'$ corrected
black hole background,
as well as from new eight-derivative terms in the action affecting
quadratic fluctuations
around the original black hole background.
Unfortunately these latter terms are not completely known at present.
In principle these terms should follow by supersymmetry from the
well known $R^4$ corrections,
but the details are not yet fully understood.

Possible corrections to the Debye mass
arise from terms quadratic in the axion coupled to either the background
curvature or the 5-form field strength.
Ref. \cite{Kehagias:1997cq} has an explicit proposal for all the
terms involving the axion and the curvature. The axion appears in
the non-perturbative contributions to the modular functions $f_k$,
but all those contributions are due to D-instantons and are
exponentially suppressed by $e^{-1/g_s} \sim e^{-4\pi\Nc/\lambda}$.
They are thus irrelevant in the large $\Nc$ limit. The other two
terms that are potentially important are $12 f_0\, R^2 DP DP$ and $6
f_2 \, R^2 (DP DP + D\bar{P} D\bar{P})$, where $P$ is proportional
to the first derivative of the axion-dilaton pair. The axion
derivative is the imaginary part of $P$. The factors $f_0$ and $f_2$
are different functions of the axion-dilaton pair, but their leading
weak coupling ({\em i.e.}, large $\Nc$) behavior is identical. In
this limit, the axion contribution miraculously cancels between the
above two terms from the proposal of Ref.~\cite{Kehagias:1997cq}.
Unfortunately the action proposed by Ref.~\cite{Kehagias:1997cq} is
not the unique $SL(2,R)$ invariant extension of the known terms
involving NSNS sector fields. In the absence of any rigorous
arguments, an explicit calculation of the coefficient of the $R^2
(\partial^2 a)^2$ term from a tree level 4-point function is needed.
This computation has been performed recently in
Ref.~\cite{Policastro:2006vt} with the result that the coefficient
is non-zero, in contradiction with the proposal of
Ref.~\cite{Kehagias:1997cq}. However the precise normalization has
not been determined, so a complete calculation of the sub-leading
correction to the Debye mass is not possible at this time. In
addition there will be non-vanishing terms involving the 5-form
field strength, which is non-zero in the AdS black-hole background,
and its derivatives. While some of these terms are known, see
Ref.~\cite{Green:2003an, deHaro:2002vk}, there may be additional
unknown terms involving the coupling of the 5-form field strength to
the axion.

We have, nevertheless, calculated the $\lambda^{-3/2}$ corrections
that would follow from the simple conjecture of
Ref.~\cite{Kehagias:1997cq}. In that case the axion fluctuation
satisfies the equation of motion \be
\partial_m\sqrt{-g} g^{mn}e^{2\phi}\partial_n \> a=0\,,
\ee
where one has to use the $\alpha'$ corrected dilaton profile
and metric in the 10$d$ Einstein frame.
Using the same shooting method as in Ref.~\cite{Csaki:1998qr},
we find that the next-to-leading order (NLO) result for the Debye mass is
\be
\mD = 3.4041 \, \big[ 1-  0.52\, \zeta(3) \,
\lambda^{-3/2} + \cdots \big] (\pi T)\,.
\label{eq:mDstrong}
\ee
As with the dilaton,
this correction leads to a smaller Debye mass
as $\lambda$ decreases.
Neglecting all further higher order corrections, this result
becomes negative at $\lambda = 0.731$.

If one assumes, for the sake of discussion, that the additional
$\lambda^{-3/2}$ corrections due to eight-derivative terms in the
corrected IIB action do not significantly change the result (\ref{eq:mDstrong}),
then it is interesting to compare the strong and weak coupling expressions
for the Debye mass.
Figure \ref{fig:weaktostrong} shows this comparison,
along with one possible smooth interpolation between weak and strong coupling
obtained by adding next-to-next-to-leading order (NNLO) terms
to both Eq.~(\ref{eq:mDweak}) and Eq.~(\ref{eq:mDstrong})
and adjusting their coefficients to make the resulting curves
tangent at a single point.%
\footnote
    {
    Specifically, $0.6 \, \lambda^{3/2} T$
    was added to the weak coupling result (\ref{eq:mDweak}),
    and $-5.0 \, \lambda^{-3}$ added inside the bracket
    of the strong coupling result (\ref {eq:mDstrong}).
    The resulting weak and strong coupling curves are
    tangent at $\lambda = 3.0$.
    \label{fn:NNLO}
    }

\begin{FIGURE}[ht]
{
\vspace*{-20pt}
\hspace*{1cm}
\includegraphics[scale=0.9,viewport=0 10 270 230,clip]{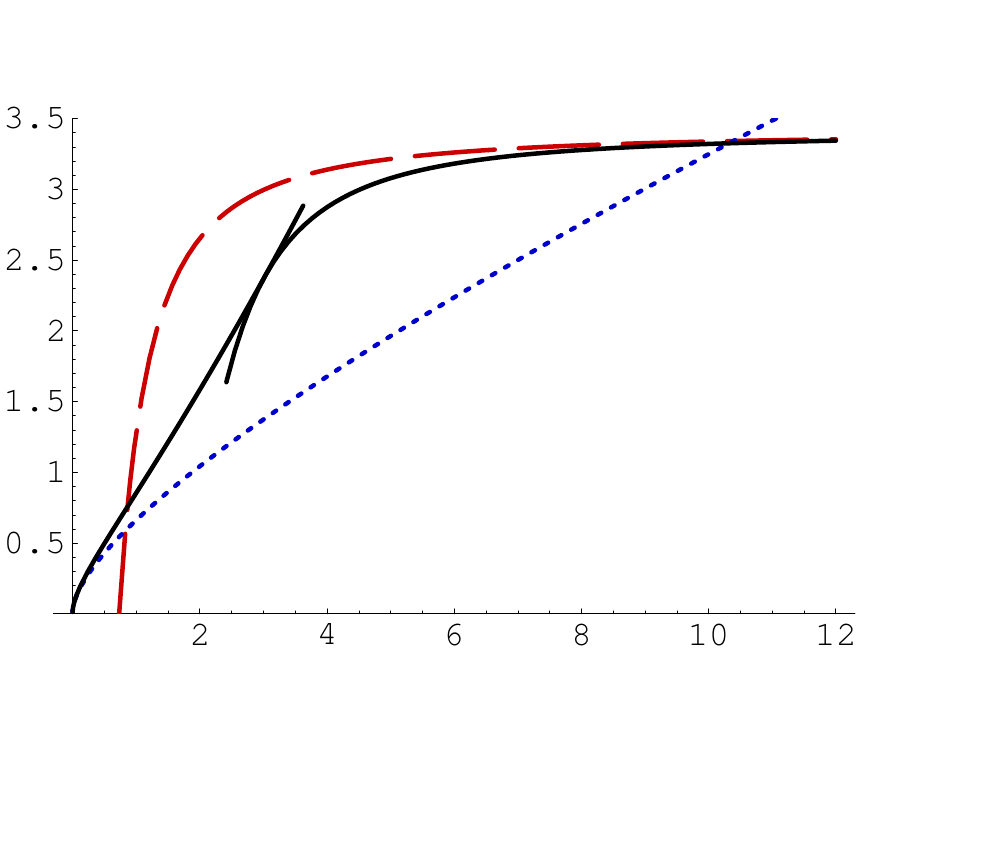}
\begin{picture}(0,0)(0,0)
\put(-10,55){$\lambda$}
\put(-250,200){$\mD/(\pi T)$}
\end{picture}
\hspace*{1cm}
\vspace*{-20pt}
\caption{\small
Debye mass in $\Nfour$ SYM as a function of $\lambda$.
The dashed (red) curve shows the NLO strong-coupling result
(\ref {eq:mDstrong}) based on the conjecture
of Ref.~\cite{Kehagias:1997cq} while the dotted (blue) curve shows the NLO
weak-coupling result (\ref {eq:mDweak}).
The solid (black) curve lying between these two NLO curves
(for intermediate couplings) shows a smooth interpolation
obtained by adding NNLO terms to both expansions,
as explained in footnote \ref{fn:NNLO},
and adjusting their coefficients to produce results which are tangent
at a single point.
\label{fig:weaktostrong}
}
}
\end{FIGURE}

This choice of interpolation is certainly not unique,
but it illustrates a reasonable choice given the available
asymptotic information.
This particular interpolation suggests that an $\Nfour$ SYM plasma
will have a Debye mass which coincides with the $T = 2\Tc$
QCD result (\ref{eq:mDcompare}) when $\lambda_{\rm SYM} \approx 2.5$.
The strong and weak coupling results would match better
({\em i.e.}, smaller NNLO terms would be needed to make them interpolate
smoothly) if the unknown additional contributions
to the Debye mass from $R^2 (\partial^2 a)^2$ terms in the effective action
lead to a larger (more negative) correction at order $\lambda^{-3/2}$.

\end{appendix}
\newpage

\bibliography{n4}
\bibliographystyle{JHEP}

\end{document}